\documentclass[11pt]{article}
\usepackage{graphicx}
\def\g{{\cal G}^{++}}
\def\G{{\cal G}^{+++}}

\def\a{\alpha}

\def\nn{\nonumber}

\newcommand{\be}{\begin{equation}}
\newcommand{\ee}{\end{equation}}
\newcommand{\bea}{\begin{eqnarray}}
\newcommand{\eea}{\end{eqnarray}}

\begin{document}
\thispagestyle{empty}
\setcounter{page}{0}
\renewcommand{\theequation}{\thesection.\arabic{equation}}

{\hfill{\tt hep-th/0505199}}

{\hfill{ULB-TH/05-10}}

\vspace{1cm}

\begin{center} {\bf \large  Dualities and signatures of  $\g$-invariant theories}

\vspace{.5cm}

Sophie de Buyl\footnote{Aspirant F.N.R.S.},  Laurent
Houart\footnote{Research Associate F.N.R.S.} and  Nassiba Tabti

\footnotesize \vspace{.5 cm}

{\em Service de Physique Th\'eorique et Math\'ematique,
Universit\'e Libre de Bruxelles,
\\ Campus Plaine C.P. 231\\ Boulevard du Triomphe, B-1050 Bruxelles,
Belgium}

\vspace{.2cm}

{\em The International Solvay Institutes, \\Campus Plaine
C.P. 231\\ Boulevard du Triomphe, B-1050 Bruxelles, Belgium}\\
{\tt sdebuyl, lhouart, ntabti@ulb.ac.be}

\end{center}

\vspace {1.5cm}

\centerline{\bf Abstract} \noindent The $\g$-content of the
formulation of gravity and M-theories as very-extended Kac-Moody
invariant theories is further analysed. The different exotic
phases of all the $\g_B$-theories, which admit exact solutions
describing intersecting branes smeared in all directions but one,
are derived. This is achieved by analysing for all $\g$ the
signatures which are related to the conventional one $(1,D-1)$ by
`dualities' generated by the Weyl reflections.
\newpage
\baselineskip18pt

\setcounter{equation}{0}
 \addtocounter{footnote}{-1}

\section{Introduction and discussion}

A theory containing gravity suitably coupled to forms and dilatons may exhibit
upon dimensional reduction down to three dimensions a simple Lie group ${\cal G}$
symmetry non-linearly realised.
The scalars of the dimensionally reduced theory live in a coset
${\cal G}/{\cal H}$ where ${\cal G}$ is in its maximally non-compact form and
${\cal H}$ is the maximal compact subgroup of ${\cal G}$. A maximally oxidised theory is such a Lagrangian
theory defined in the highest possible space-time dimension $D$ namely a theory which is itself not
obtained by dimensional reduction. These maximally oxidised actions have been constructed for all   $\cal
G$ \cite{cjlp} and they include in particular pure gravity in $D$
dimensions and the low energy effective actions of the bosonic string and
of  M-theory.
\begin{figure}[h]
   \centering
   \includegraphics[width=12cm]{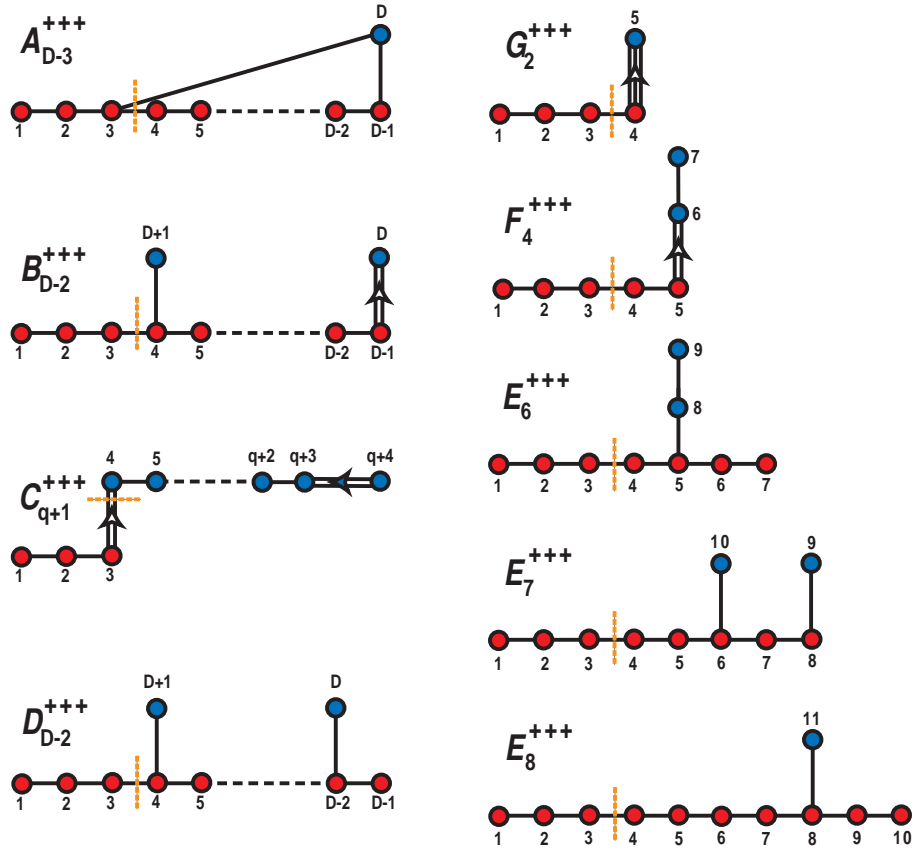}
 \caption { \small The
nodes labelled 1,2,3 define the Kac-Moody extensions of the  Lie
algebras. The horizontal line
starting at 1 defines the `gravity line', which is the
Dynkin diagram  of a
$A_{D-1}$ subalgebra.}
   \label{first}
\end{figure}

It has been conjectured that these theories, or some extensions of them,
possess the much larger very-extended Kac-Moody symmetry
$\G$. $\G$ algebras are defined by the Dynkin diagrams depicted
in Fig.1,  obtained from those of $\cal G$ by adding three
nodes~\cite{ogw}. One first adds the affine node, labelled 3 in the
figure, then a second node, 2,  connected to it by a single line and
defining the overextended ${\cal G}^{++}$ algebra\footnote{In the context of dimensional reduction,   the
appearance of
$E_8^{++}=E_{10}$ in one dimension has been first conjectured by B. Julia
\cite{julia84}.},   then  a third
one, 1, connected  by a single line to the overextended node. Such
$\G$ symmetries were first conjectured in the aforementioned
particular cases
\cite{west01,lw} and the extension to all
$\G$ was proposed in
\cite{ehtw}. In a different development, the study of the properties
of cosmological solutions in the vicinity of a space-like singularity,
known as cosmological billiards
\cite{damourhn00},  revealed an  overextended symmetry ${\cal G}^{++}$
for all maximally oxidised theories
\cite{damourh00, damourbhs02}.

The possible existence of this Kac-Moody symmetry $\G$ motivated
the construction of a Lagrangian formulation explicitly invariant
under $\G$  \cite{eh}. The action $S_{\G}$ is defined in a
reparametrisation invariant way on a world-line, a priori
unrelated to space-time, in terms of fields $\phi(\xi)$ living in
a coset $\G/K^{+++}$ where $\xi$ spans the world-line. A level
decomposition of $\G$ with respect to the subalgebra $A_{D-1}$ of
its gravity line (see Fig. 1) is performed where $D$ is
identified to the space-time dimension\footnote{ Level expansions
of very-extended algebras in terms of the subalgebra $A_{D-1}$
have been considered in \cite{west02, nifi, weke}.}. The
subalgebra  $K^{+++}$ is invariant under a `temporal' involution
which ensures that the action is $SO(1,D-1)$ invariant at each
level where the index $1$ of $A_{D-1}$ is identified to a time
coordinate.

The $\g$ content of the $\G$ -invariant actions $S_{{\cal
G}^{+++}}$ has been analysed in reference \cite{ehh} where it was
shown that two distinct actions invariant under the overextended
Kac-Moody algebra $\g$ exist. The first one $S_{\g_C}$ is
constructed from $S_{\G}$ by performing a truncation putting
consistently to zero some fields. The corresponding $\g$  algebra
is obtained from $\G$ by deleting the node labelled 1 from the
Dynkin diagram of $\G$ depicted in Fig. 1. This theory carries a
Euclidean signature and is the generalisation to all $\g$ of
the $E_8^{++}=E_{10}$ invariant action of reference
\cite{damourhn02} proposed in the context of M-theory and
cosmological billiards. The parameter $\xi$ is then identified
with the time coordinate and the action restricted to a defined
number of lowest levels is equal to the
corresponding maximally oxidised theory in which the fields depend
only  on this time coordinate. A second $\g$-invariant action
$S_{\g_B}$ is obtained from $S_{\G}$ by performing the same
consistent truncation {\it after} conjugation by the Weyl
reflection $W_{\a_1}$ in $\G$ where $W_{\a_1}$ is the Weyl
reflection in the hyperplane perpendicular to  the simple root
$\a_1$ corresponding to the node 1 of figure 1. The
non-commutativity of the temporal involution with the Weyl
reflection \cite{keu1,keu2} implies that this second action is
inequivalent to the first one. In $S_{\g_B}$, $\xi$ is identified
with a space-like direction and the theory admits exact solutions
which are identical to those of the corresponding maximally
oxidised theory describing intersecting extremal brane
configurations smeared in all directions but one
\cite{ehh,eh,eh2}. Furthermore the intersection rules \cite{aeh}
are neatly encoded in the $\g$ algebra through orthogonality
conditions between the real positive roots corresponding to the
branes in the configuration \cite{eh2}.

The Weyl reflections of $\G$ generated by roots not belonging to the gravity line are also Weyl
reflection of $\g$, their actions on $S_{\g_B}$ are thus well defined. These reflections yield actions
with different global signatures which are related by field redefinitions \cite{ehh}. This equivalence
realises in the action formalism the general analysis of Keurentjes \cite{keu1,keu2}.

The precise analysis of the different possible signatures has been performed for $\g_B=E_8^{++}=E_{10}$.
In this case the corresponding maximally oxidised theory is the bosonic sector of the  low effective action of M-theory.
For $E_8^{++}$ the Weyl reflection $W_{\a_{11}}$ generated by the simple root $\a_{11}$ (see Fig.
1) corresponds to a double T-duality in the directions 9 and 10 followed by an exchange of the
two radii \cite{ weylt,banks,ehtw}. The signatures found in the analysis of references \cite{keu1,keu2} and in the context of $S_{\g_B}$ in \cite{ehh} match perfectly with the signature changing dualities and the exotic
phases of M-theories discussed in \cite{hull1,hull2}.

The action of Weyl reflections generated by simple roots not
belonging to the gravity line on the exact extremal brane
solutions has been studied for all $\G$ -theory constructed with
the temporal involution selecting the index 1 as a time coordinate
\cite{eh}. The existence of Weyl orbits of extremal brane
solutions similar to the U-duality orbits existing in M-theory
strongly suggests a general group-theoretical origin of
`dualities' for all $\G$ -theories transcending string theories and
supersymmetry.

In this context it is certainly interesting to extend to all
$\G$-theories the analysis of signature changing Weyl reflections.
This is the purpose of the present work.   We find for all the
$\g_B$-theories all the possible signature $(T,S)$, where $T$
(resp. $S$) is the number of time-like (resp. space-like)
directions, related by Weyl reflections of $\g$ to the signature
$(1,D-1)$ associated to the theory corresponding to the
traditional maximally oxidised theories. Along with the different
signatures the signs of the kinetic terms of the relevant fields
are also discussed. We start the analysis with $A_{D-3}^{++}$
corresponding to pure gravity in $D$ dimensions then we extend the
analysis to the other $\g$, first to the simply laced ones and then to the
non-simply laced ones\footnote{Arjan Keurentjes informed us that he obtained independently similar 
results to be released soon.}. Each $\g$ algebra contains a $A_{D-3}^{++}$
subalgebra, the signatures of $\g$ should thus includes the one of
$A_{D-3}^{++}$. This is indeed the case, but some $\g$ will
contain additional signatures. If one want to restrict our focus
on  string theory, the special cases of $D_{24}^{++}$ and
$B_{8}^{++}$ are interesting, the former being related to the
low-energy effective action of the bosonic string (without
tachyon) and the latter being related to the low-energy effective
action of the heterotic string (restricted to one gauge field).
The existence of signature changing dualities  are related to the
magnetic roots and suggests that these transformations correspond
to a generalisation of the S-duality existing in these two
theories \cite{sen}.

 The  approach  based on Kac-Moody algebras constitutes certainly a very-exciting and innovative attempt to understand
 gravitational theories encompassing string theories. It is certainly worthwhile to pursue this route by analysing further the structure of these $\g \subset \G$ -theories and to try to understand if it could lead to a completely
 new formulation of gravitational interactions where the structure of space-time is hidden somewhere in these huge algebras \cite{damourhn02,ehh,dani} or even huger ones \cite{e12}.

The paper is organised as follows. In section 2, we recall the
construction of $S_{\G}$-invariant actions. In section 3, we
review the non-commutativity of the temporal involution with the
Weyl reflections and develop the notation and tools necessary to
discuss signature changes in our setting.
 In section 4, we review briefly the two different $\g$-theories obtained by truncation of
 the $S_{\G}$ action. Finally in section 5, we derive all the possible signatures for all the
 $\g_B$-theories.

\section{The $\G$-invariant theories}

In this section, we recall the construction of the
$\G$-invariant theories
\cite{eh}.
Actions $S_{\G}$ invariant under non-linear
transformations of
$\G$ are constructed recursively from a level
decomposition with
respect to a subalgebra
$A_{D-1}$ where $D$ is interpreted as the space-time dimension.
Each $\G$ contains indeed a   subalgebra $GL(D)$ such that $SL(D) (=A_{D-1})
\subset GL(D) \subset
\G$. The
action is defined in a reparametrisation invariant way  on a
world-line, a priori unrelated to space-time, in terms of fields
$\varphi(\xi)$ where $\xi$ spans the world-line. The fields $\varphi(\xi)$
live in a coset space
$\G/K^{+++}$ where the subalgebra $K^{+++}$ is invariant under a
`temporal involution'
 preserving at each level a Lorentz algebra
$SO(1,D-1)= A_{D-1} \cap K^{+++}$.

The generators of the $GL(D)$ subalgebra are taken to be
$K^a_{~b}\ (a,b=1,2,\ldots ,D)$   with commutation relations
\begin{equation}
\label{Kcom} [K^a_{~b},K^c_{~d}]   =\delta^c_b
K^a_{~d}-\delta^a_dK^c_{~b}\,  .
\end{equation}  The $K^a_{~b}$ along with  abelian generators $R_u \,( u=1 \dots q)$,
which are present  when the corresponding maximally oxidised action
$S_{\cal G}$ has $q$ dilatons\footnote{All the maximally oxidised  theories have at most one dilaton except
the $C_{q+1}$-series characterised by $q$ dilatons. }, are the level zero
generators. The step operators of level greater than zero are tensors
$R^{\quad c_1\dots c_r}_{ d_1\dots d_s}$ of the
$A_{D-1}$ subalgebra. Each  tensor forms an  irreducible representation of $A_{D-1}$ characterised
by some Dynkin labels.  In principle it is possible to determine the irreducible representations
present at each level \cite{nifi,weke}.
 The lowest levels contain antisymmetric tensor step  operators
$R^{a_1a_2 \dots a_r}$ associated to electric and magnetic
roots arising from the dimensional reduction of field strength forms in $S_{\cal
G}$. They satisfy the tensor and scaling relations
\begin{eqnarray}
\label{root} &&[K^a_{~b},R^{a_1\dots a_r}]   =\delta^{a_1}_b R^{aa_2
\dots a_r} +\dots +
\delta^{a_r}_b R^{a_1 \dots a_{r-1}a}\, ,\\
\label{root2} && [R, R^{a_1\dots a_r}
] =   -\frac{\varepsilon_A
a_A}{2}\,  R^{a_1\dots a_r}\, ,
\end{eqnarray} where $a_A$ is the dilaton coupling constant to the
 field strength form   and
$\varepsilon_A$ is $+1\, (-1)$ for an
electric (magnetic) root \cite{ehtw}. The generators obey the invariant
scalar product relations
\begin{eqnarray}
\label{killing}
&&\langle K_{~a}^a K_{~b}^b\rangle =G_{ab}\, ,\quad \langle
K^b_{~a}K_{~c}^d\rangle=
\delta_c^b\delta_a^d \ a\neq b\, , \quad\langle R R\rangle
=\frac{1}{2}
\,,\\
\label{step}
&&\langle
R^{\quad a_1\dots a_r}_{ b_1\dots b_s} , \bar  R_{ d_1\dots   d_r}^{\quad
c_1\dots c_s}\rangle
=\delta^{c_1}_{b_1}\dots\delta^{c_s}_{b_s}\delta^{a_1}_{d_1}\dots\delta^
{a_r}_{d_r}\,.
\end{eqnarray}
Here $G= I_D -
\frac{1}{2}\Xi_D$ where $\Xi_D$ is a $D$-dimensional matrix with all
entries   equal to unity and   $\bar  R_{ d_1\dots   d_r}^{\quad
c_1\dots c_s}$ designates the negative step operator conjugate to $ R^{\quad
d_1\dots   d_r}_{ c_1\dots c_s}$.

The temporal involution $\Omega_1$ generalises the Chevalley
involution to allow identification of the index 1 to a time coordinate in
$SO(1,D-1)$. It is  defined by
\begin{equation}
\label{map} K^a_{~b}\stackrel{\Omega_1}{\mapsto}
-\epsilon_a\epsilon_b K^b_{~a}\quad R\stackrel{\Omega_1}{\mapsto} -R\quad
, \quad R^{\quad c_1\dots c_r}_{ d_1\dots d_s}
\stackrel{\Omega_1}{\mapsto}
-\epsilon_{c_1}\dots\epsilon_{c_r}\epsilon_{d_1}\dots\epsilon_{d_s}
  \bar R_{ c_1\dots c_r}^{\quad d_1\dots d_s}\, ,
\end{equation}
 with $\epsilon_a =-1$ if $a=1$ and
$\epsilon_a=+1$ otherwise. It leaves invariant a subalgebra $K^{+++}$ of
$\G$.

 The fields $\varphi(\xi)$ living in the coset space ${\G}/K^{+++}
$ parametrise the Borel group  built out of Cartan and positive
step operators in
$\G$. Its elements $\cal V$  are written as
\begin{equation}
\label{positive} {\cal V(\xi)}= \exp (\sum_{a\ge b}
h_b^{~a}(\xi)K^b_{~a} -\sum_{u=1}^q
\phi^u(\xi) R_u) \exp (\sum
\frac{1}{r!s!} A^{\quad a_1\dots a_r}_{ b_1\dots b_s}(\xi) R_{
a_1\dots   a_r}^{\quad b_1\dots b_s} +\cdots)\, ,
\end{equation}
 where the first exponential contains only  level zero  operators and
the second one the positive step operators of levels strictly
greater than zero. Defining
\begin{equation}
\label{sym}
dv(\xi)= d{\cal V} {\cal V}^{-1}\quad d\tilde
v(\xi)=   -\Omega_1
\, dv(\xi)
\qquad\quad dv_{sym}=\frac{1}{2} (dv+d\tilde v)\, ,
\end{equation}
one obtains, in terms of the
$\xi$-dependent fields, an  action $S_{\cal \G}$ invariant under
global $\G$ transformations, defined on  the
coset ${\G}/K^{+++}$
\begin{equation}
\label{actionG} S_{\cal \G}=\int d\xi  \frac{1}{n(\xi)}\langle
(\frac{dv_{sym}(\xi)}{d\xi})^2\rangle\, ,
\end{equation} where
$n(\xi)$ is an arbitrary lapse function ensuring reparametrisation
invariance on the world-line.

Writing
\begin{equation}
\label{full} S_{{\cal G}^{+++}} =S_{{\cal G}^{+++}}^{(0)}+\sum_A
S_{{\cal G}^{+++}}^{(A)}\, ,
\end{equation} where $S_{{\cal G}^{+++}}^{(0)}$ contains all level
zero contributions, one obtains
\begin{equation}
\label{fullzero} S_{{\cal G}^{+++}}^{(0)} =\frac{1}{2}\int d\xi
\frac{1}{n(\xi)}\left[\frac{1}{2}(g^{\mu\nu}g^{\sigma\tau}-
\frac{1}{2}g^{\mu\sigma}g^{\nu\tau})\frac{dg_{\mu\sigma}}{d\xi}
\frac{dg_{\nu\tau}}{d\xi}+\sum_{u=1}^q
\frac{d\phi^u}{d\xi}\frac{d\phi^u}{d\xi}\right],
\end{equation}
\begin{equation}
\label{fulla} S_{{\cal G}^{+++}}^{(A)}=\frac{1}{2 r! s!}\int d\xi
\frac{ e^{- 2\lambda
\phi}}{n(\xi)}\left[
\frac{DA_{\mu_1\dots \mu_r}^{\quad \nu_1\dots
\nu_s}}{d\xi} g^{\mu_1{\mu}^\prime_1}...\,
g^{\mu_r{\mu}^\prime_r}g_{\nu_1{\nu}^\prime_1}...\,
g_{\nu_s{\nu}^\prime_s}
\frac{DA_{{\mu}^\prime_1\dots {\mu}^\prime_r}^{\quad
{\nu}^\prime_1\dots {\nu}^\prime_s}}{d\xi}\right].
\end{equation}  The $\xi$-dependent fields $g_{\mu\nu}$ are defined as
$g_{\mu\nu} =e_\mu^{~a}e_\nu^{~b}\eta_{ab}$ where $e_\mu^{~a}=(e^{-h(\xi)})_\mu^{~a}$. The appearance of the
Lorentz metric $\eta_{ab}$ with $\eta_{11}=-1$ is a consequence of the
temporal involution $\Omega_1$. The metric $g_{\mu\nu}$ allows a
switch from  the Lorentz  indices  $(a,b)$ of the fields appearing  in
Eq.(\ref{positive}) to
$GL(D)$ indices $(\mu,\nu)$.  $D/D\xi$ is a  covariant derivative generalising
$d/d\xi$ through  non-linear terms arising from  non-vanishing
commutators  between  positive  step operators and  $\lambda$ is the
generalisation of the scale parameter
$-\varepsilon_A a_A/2$ to all roots.

\section{Weyl reflections and the temporal involution}

The non-commutativity
of the temporal involution with the Weyl reflections implies that different space-time
signatures are related between themselves \cite{keu1,keu2}. In this section, we recall
the basic facts,  we set up
the notations and recollect some tools and formulae
necessary for the general discussion of signature changes presented in section 5
in the setting of our construction of $\g$-actions \cite{ehh}.

A  Weyl transformation $W$  can be expressed as a conjugation by a group
element
$U_W$ of $\G$. We define the involution $\Omega^\prime$ operating on
the conjugate elements by
\begin{equation}
\label{newinvolve}
U_W\, \Omega T\, U^{-1}_W=\Omega^\prime \, U_W  T U^{-1}_W\, ,
\end{equation}
where $T$ is any generator of $\G$.

We first recall the effect of the Weyl reflection $W_{\alpha_1}$
generated by the simple root $\a_1$ (see Fig. 1). One gets

\begin{eqnarray}
\label{permute}
&&U_1\, \Omega K^2_{\ 1} \, U^{-1}_1= \rho K^2_{\ 1}= \rho\Omega^\prime
 K^1_{\ 2}\nonumber\, ,\\
&&U_1\, \Omega K^1_{\ 3} \, U^{-1}_1= \sigma K^3_{\ 2}=
\sigma\Omega^\prime
 K^2_{\ 3} \, ,\\
&&U_1\, \Omega K^i_{\ i +1} \, U^{-1}_1= -\tau K^{i+1}_{\ \, i}=
\tau\Omega^\prime  K^i_{\ i +1}\quad i >2\, .\nonumber
\end{eqnarray}
Here $\rho,\sigma,\tau$ are plus or minus signs which may arise
as step operators are representations of the Weyl group
up to signs. Eq.(\ref{permute}) illustrate the general result
that such signs always cancel in the determination of
$\Omega^\prime$ . The content of Eq.(\ref{permute}) is
represented in Table 1. The signs below the generators of the gravity
line indicate the sign in front of the
 negative step operator obtained by the involution: a
minus sign is in agreement with the conventional Chevalley involution
and indicates that the indices in
$K^a_{\ a +1}$ are both either space or time indices while a plus sign
indicates that one index must be time and the other  space.
\begin{table}[h]
\caption{\small Involution switch from $\Omega$ to
$\Omega^\prime$ due to the Weyl reflection $W_{\alpha_1}$}
\begin{center}
\begin{tabular}{|c|ccccc|c|}
\hline
gravity line&$K^1_{\ 2}$&$K^2_{\ 3}$&$K^3_{\ 4}$&$\cdots$&$K^{D-1}_{\
D}$&time coordinate\\
\hline\hline
$\Omega$&$+$&$-$&$-$&$-$&$-$&1\\
\hline$\,\Omega^\prime$&$+$&$+$&$-$&$-$&$-$&2\\
\hline
\end{tabular}
\end{center}
\label{gravity}
\end{table}

\noindent
If we choose the description which leaves
unaffected coordinates attached to planes invariant under the Weyl
transformation, Table 1 shows that
the  time coordinate must be identified with 2.
The generic Weyl reflection $W_{\a_a}$ generated by  $\a_a$ a simple root of the gravity line
exchanges the index $a$ and $a+1$ along with the space-time nature of the corresponding coordinates.

Weyl reflections generated by simple roots not belonging to the gravity line
relate step operators of different levels. As a consequence \cite{keu1,keu2}, these
 may potentially induce changes of signature far less trivial than the simple exchange of the index
 identifying the time coordinate . These changes have been studied from the algebraic point
 of view in great details for $E_{11}$  (and more generally for $E_n$) in
 \cite{keu1,keu2}\footnote{Some algebraic considerations in this context for others groups $\cal G$ are presented in the Appendix of the second reference.}.

 In order to address the action of an  involution on a generic step operator
 $R^{a_1 \dots a_r}$ of level greater than zero in a given irrreducible representation of $A_{D-1}$, we introduce some notations.
 First, given an involution $\tilde \Omega$, one defines
 $\mathrm{sign}(\tilde \Omega X)$ for any given positive step operator  $X$  in  the following way
 \begin{equation}
 \tilde \Omega \, X \equiv \mathrm{sign}(\tilde \Omega X) \, {\bar X},
 \label{sidef}
 \end{equation}
 where $\bar X$ designates the negative step operator conjugate to $X$.
Second, we also introduce a sign associated to a given  positive step operator of level greater than zero\footnote{As far as the action of the involution is concerned the symmetry properties of a step operator given by its Dynkin labels do not play any role.}
 $R^{a_1 \dots a_r}$
in the following way

 \begin{eqnarray}
\label{signa}
&+ \quad : &\mathrm{sign}(\tilde \Omega R^{a_1 \dots a_r})= -\epsilon_{a_1}\dots\epsilon_{a_r}  \\
\label{signb}
&- \quad : &\mathrm{sign}(\tilde \Omega R^{a_1 \dots a_r})=+\epsilon_{a_1}\dots\epsilon_{a_r},
\end{eqnarray}
where $\epsilon_a=-1$ if $a$ is a timelike index and
$\epsilon_a=+1$ if $a$ is a space-like index, the space-time
nature of the coordinate labelled by the index $a$ being defined
by the action of $\tilde \Omega$ on the $K^a_{\, b}$. The $+$ sign
defined in Eq.(\ref{signa}) will lead to the positive kinetic
energy term for the corresponding field in the action while
the $-$ sign defined in Eq.(\ref{signb}) will lead to a negative
kinetic energy term.

Finally, if we perform a Weyl reflection $W_Y$ generated by a
simple root not belonging to the gravity line and associated to a
step operator $Y$ Eq.(\ref{newinvolve}) gives
\begin{equation}
\label{kisi}
\mathrm{sign}(\Omega Y) = \mathrm{sign}(\Omega^\prime Y),
\end{equation}
because $\mathrm{sign}(\tilde \Omega Y)= \mathrm{sign}(\tilde \Omega {\bar Y})$ where
$\bar Y$ is the negative step operator conjugate to $Y$.

\section{From  the $\G$-theory  to  the $\g_C$ and $\g_B$-theories }

It has been recently shown \cite{ehh} that for each very-extended algebra $\G$, the $\G$- invariant theory encompasses
two distinct theories invariant under the overextended Kac-Moody subalgebra $\g$.
The  $\g_C$- invariant action $S_{{\cal G}_C^{++}}$ describes a motion in a coset    $\g/K^{++}_{(+)}$ and carries a Euclidean signature while the second theory described by a different embedding of $\g$ in
$\G$, referred as $\g_B$,  describes a motion in a different coset  $\g/K^{++}_{(-)}$. In contradistinction  with
the $\g_C$ case, the $\g_B$-theory
carries various Lorentzian signatures which are revealed through various equivalent
formulation related by Weyl transformations. We now recall  the construction of these two theories.

\subsection{ The $\g_C$-theory}
Consider the overextended algebra $\g_C$ obtained from the very-extended
algebra $\G$  by deleting the  node labelled 1 from the Dynkin
diagrams of $\G$ depicted in Fig.1.
The action $S_{\g_C}$ describing the $\g_C$ theory is obtained from $\G$ by performing the following
consistent truncation. One puts to zero in  the coset representative Eq.(\ref{positive}) the
field multiplying the Chevalley generator
$H_1=  K_{~1}^1- K_{~2}^2$ and all the
fields multiplying the positive step operators associated to  roots whose
decomposition in terms of simple roots contains  the deleted root
$\alpha_1$. Performing this truncation one obtains the following action
\begin{equation}
\label{fullp} S_{\g_C} = S_{\g_C}^{(0)}+\sum_B
S_{\g_C}^{(B)}\, ,
\end{equation} where
\begin{eqnarray}
\label{fullop} S_{\g_C}^{(0)}& =&\frac{1}{2}\int dt
\frac{1}{n(t)}\left[\frac{1}{2}(g^{\hat{\mu}\hat{\nu}}g^{\hat{\sigma}\hat{\tau}}-
g^{\hat{\mu}\hat{\sigma}}g^{\hat{\nu}\hat{\tau}})\frac{dg_{\hat{\mu}
\hat{\sigma}}}{dt}
\frac{dg_{\hat{\nu}\hat{\tau}}}{dt}+
\frac{d\phi}{dt}\frac{d\phi}{dt}\right] ,\\
\label{fullc} S_{\g_C}^{(B)}&=&\frac{1}{2 r! s!}\int dt
\frac{ e^{- 2\lambda
\phi}}{n(t)}\left[
\frac{DB_{\hat{\mu}_1\dots \hat{\mu}_r}^{\quad \hat{\nu}_1 \dots
\hat{\nu}_s}}{dt} g^{\hat{\mu}_1{\hat{\mu}}^\prime_1}...\,
g^{\hat{\mu}_r{\hat{\mu}}^\prime_r}g_{\hat{\nu}_1{\hat{\nu}}^\prime_1}
...\ ,    g_{\hat{\nu}_s{\hat{\nu}}^\prime_s}
\frac{DB_{{\hat{\mu}}^\prime_1\dots {\hat{\mu}}^\prime_r}^{\quad
{\hat{\nu}}^\prime_1\dots {\hat{\nu}}^\prime_s}}{dt}\right] .
\end{eqnarray} where  the
hatted indices $\hat\mu , (\mu=2,\dots D)$ have been introduced, the remaining $A$-fields have been denoted by $B$ and  $\xi$  has been renamed  as $t$.

This theory describes a motion on the coset
$\g/K^{++}_{(+)}$ where  $K^{++}_{(+)}$ is the subalgebra of $\g$
invariant under the Chevalley involution. This `cosmological' action $S_{\g_C}$ generalises to
all $\g$  the $E_{10}\, (\equiv E_8^{++})$ action of reference \cite{damourhn02, damourn04} proposed in the context of M-theory and cosmological billiards. Since  $K^{++}_{(+)}$ is defined by the Chevalley
involution which commutes with the Weyl reflection of $\g$ this coset admits only the Euclidean signature.

\subsection{ The $\g_B$-theory}

The $\g_B$-theory is obtained by performing the same truncation
as the one of section 4.1, namely we equate as before to zero
all the fields in the $\G$-invariant action Eq.(\ref{full}) which
multiply generators  involving the root $\alpha_1$, but this
truncation is performed {\em after} the $\G$ Weyl transformation Eq.(\ref{permute})
which
transmutes the time index 1 to a space index. This gives an action $S_{\g_B}$ which is
formally identical to the one given by Eqs.(\ref{fullp}),
(\ref{fullop}) and (\ref{fullc}) but with a Lorentz signature for the
metric, which in the flat coordinates amounts to a negative sign for the
Lorentz metric component $\eta_{22}$, and with $\xi$ identified to the missing
space coordinate  instead of $t$.
This theory admits exact solutions identical to intersecting extremal brane solutions of the corresponding maximally oxidised theory smeared in
all direction but one \cite{ehh}. These solutions  provide a laboratory to understand the significance of the higher level fields and to check whether or not the Kac-Moody theory can
described uncompactified theories \cite{damourhn02,ehh,dani}.

The $S_{\g_B}$ action is thus characterised by a signature $(1,D-2,+)$
where  the sign $+$ means that Eq.(\ref{signa}) is fulfilled for all the simple positive step operators
implying that all the kinetic energy terms in the action are positive.
The theory describes a motion on the coset
$\g/K^{++}_{(-)}$.  $K^{++}_{(-)}$ is the subalgebra of $\g$
invariant under the  time involution $\Omega_2$ defined as in
Eq.(\ref{map}) with 2 as the time coordinate and restricted to $\g$.
As this involution will not generically commute with Weyl reflections, the same coset can be
described by actions $S_{{\cal G}{}_{(i{}_1i{}_2\dots i{}_t)}^{++}}^{(T,S,\varepsilon)}$, where
the global signature is $(T,S,\varepsilon)$ with
$\varepsilon$ denoting  a set of  signs, one for each
simple step operator which does not belong to the gravity line, defined by Eqs.(\ref{signa}) and (\ref{signb}), and $i{}_1i{}_2\dots i{}_t$ are the time indices. The equivalence of the different actions has been
shown by deriving differential equations relating the fields parametrising the different coset
representatives \cite{ehh} and in the special case of $\g_B=E_8^{++}$ all the signatures in the orbits of
$(1,D-2,+)$ has been found and agreed with \cite{keu1,keu2} and \cite{hull1,hull2}.
In the next section we derive all the signatures in the orbit of $(1,D-2,\{ \epsilon =+\})$ for all $\g_B$-theories.

\section{The signatures of $\g_B$-theories}
We will characterise the different signatures of $\g_B$   in
terms of $\G$, namely  we will determine all the signatures in the
Weyl orbit of $(1,D-1,\{ \epsilon =+\})$ with the index 1 fixed to be  a
space-like coordinate. We first discuss in detail the
$\g_B=A_{D-3}^{++}  \subset A_{D-3}^{+++}$ case. The other $\G$
contain as a subalgebra $A_{D-3}^{+++}$ (there is always a
graviphoton present at some level) consequently the signatures of
all $\g_B$ include at least those of $A_{D-3}^{++}$.

\subsection{$A_{D-3}^{+++}$}

\subsubsection{$D>5$}

Our purpose is to determine all $S_{{A}{}_{(i{}_1i{}_2\dots
i{}_t)}^{++}}^{(T,S,\varepsilon)}$ equivalent to $S_{A^{++}_B}$,
i.e. all $\Omega^\prime$ related to $\Omega_2$ via a Weyl
reflection of $A_{D-3}^{++}$ (see Eq.(\ref{newinvolve})). As
explain above, the only Weyl reflections changing  the signature
in a non-trivial way  are the ones generated by simple roots not
belonging to the gravity line. Here there is only one such a
simple root , namely $\a_D$ (see Fig. 1). The  Weyl reflection
$W_{\a_D}$ exchanges the following roots,
\begin{eqnarray}
\label{ex1}
\alpha_{D-1} & \leftrightarrow & \alpha_{D-1} + \a_D \\
 \label{ex2}
 \a_{3} &  \leftrightarrow &  \a_3 + \a_{D}.
 \end{eqnarray}
One can express this Weyl reflection as a conjugaison by a group
element $U_{W_{\a_D}}$ of $A_{D-3}^{+++}$. The non-trivial action
of $U_{W_{\a_D}}$, on the step operators is given by,

\begin{eqnarray}
\label{exc1}
K^{D-1}{}_D & \leftrightarrow  & \sigma  R^{4...D,D-1} \\
\label{exc2}
K^3{}_4 &\leftrightarrow &  \rho R^{35...D,D},
 \end{eqnarray}
where $\sigma$ and $\rho$ are +1 or -1 and the tensor
$R^{a_4...a_{D},a_{D+1}}$ is in the representation\footnote{Here
we adopt the following convention : the   Dynkin labels of the
$A_{D-1}$ representations are labelled  from right to left when
compared with the labelling of the Dynkin diagram of Fig.1. For
instance the last label on the right refers to the fundamental
weight associated with the root   labelled 1 in Fig.1. In
\cite{weke}, the opposite convention is used.} $[1,0,\dots 1,0,0]$
of $A_{D-1}$ that occurs at level one \cite{weke}. $ R^{4...D,D-1}$ is the generator of
$A_{D-3}^{+++}$ associated to the root $\a_{D-1} + \a_D$ and  $
R^{35...D,D}$ the one associated to $ \a_3 + \a_{D}$.

In order to obtain  a change of  signature, we need generically  $\Omega
K^3{}_4 \neq \Omega' K^3{}_4$ and/or $\Omega K^{D-1}{}_D \neq
\Omega' K^{D-1}{}_D$. Using Eqs.(\ref{exc1}), (\ref{exc2}) and Eq.(\ref{sidef}), 
these conditions are equivalent to
$\mathrm{sign} ( \Omega K^3{}_4 ) \neq \mathrm{sign} ( \Omega
R^{35...D,D} ) $ and/or $ \mathrm{sign} ( \Omega K^{D-1}{}_D )=
\mathrm{sign} ( \Omega R^{4...D,D-1} )$. The following
equalities,

\bea \label{signcond} \mathrm{sign} (\Omega  R^{4...D,D} ) &=&-
\mathrm{sign}(\Omega R^{35...D,D}).\mathrm{sign}(\Omega K^3{}_4) \nn \\
&=& -\mathrm{sign} (\Omega R^{4...D,D-1} ).\mathrm{sign}(\Omega
K^{D-1}{}_D),
\eea imply that we have only $\Omega
K^3{}_4 \neq \Omega' K^3{}_4$ \textit{and} $\Omega K^{D-1}{}_D
\neq \Omega' K^{D-1}{}_D$. These inequalities lead to four different possibilities
summarised in Table \ref{con1}.

\begin{table}[h]
\caption{\small  Conditions on $\Omega$'s leading to non-trivial signature
changes under the Weyl reflection $W_{\alpha_D}$}
\begin{center}
\begin{tabular}{|c|c|c|c|}
\hline  & $ \mathrm{sign}(\Omega K^3{}_4$) & $\mathrm{sign}(\Omega
R^{35...D,D})$ & $\mathrm{sign}(\Omega
K^{D-1}{}_D) $  \\
\hline
 a. & + & - & +   \\
 b. & + & - & -  \\
 c. & - & + & +   \\
 d. & - & + & -   \\
\hline
\end{tabular}
\end{center}
\label{con1}
\end{table}
Note that  by  Eq. (\ref{signcond}) the sign of $\Omega R^{4...D,D-1}$ is
deduced from the signs of $ \Omega K^3{}_4$, $\Omega R^{35...D,D}$
and $\Omega K^{D-1}{}_D $. All the signatures in the orbit of $(1,D-1,+)$, where $+$ is the sign
associated to the generatorÊ $R^{a_4...a_{D},a_{D+1}}$ defined
by Eq.(\ref{signa}), are now derived.

\begin{itemize}
\item \textbf{Step 1}: Let us first consider a $\Omega$ characterised by a signature $(T,S,+)$  where $T$ is
odd. We consider this set of signatures because it contains our starting point
$(1,D-1,+)$ and also because other signatures of this type will be useful for the
recurrence, i.e. the signature obtained from $(1,D-1,+)$
will lead in some cases to new signatures of the general type
$(T,S,+)$ where $T$ is odd. We analyse the different possibilities of Table \ref{con1}.

\begin{itemize}
\item a. The coordinates $3$ and $4$ are of  different nature as well as the coordinates $D-1$ and $D$. Moreover they are an even
number of time coordinates in the set $\{ 3,5,...,D-1\}$ as a direct consequence of the sign $-$ in the second column of
Table \ref{con1} and the fact that $R^{a_4...a_{D},a_{D+1}}$ satisfy Eq.(\ref{signa}) . So they are an odd number of time coordinates in
the complementary subset $\{1,2,4,D\}$, i.e 1 or 3. These
conditions lead to the possibilities given in Table
\ref{gravity1}.

\begin{table}[h]
\caption{\small $\Omega$'s leading to  signature changes under the
Weyl reflection $W_{\alpha_D}$.  Space-like (resp. time-like)
coordinate are denoted by s (resp. t)}
\begin{center}
\begin{tabular}{|c|ccc|ccc|c|}
\hline &       1 & 2 & 3 & 4 & ... & D-1 & D \\
\hline a.1. &  s & t & t & s & ... & t & s \\
       a.2. &  s & s & s & t & ... & t & s \\
       a.3. &  s & s & t & s & ... & s & t \\
       a.4. &  s & t & s & t & ... & s & t \\
\hline
 \end{tabular}
\end{center}
 \label{gravity1}
\end{table}
 The nature of the coordinates 1, 2, 3 and $D$ (resp.
4,...,$D-1$) does not (resp. does) change under the action of the
Weyl reflection generated by $\alpha_{D}$ on $\Omega$'s satisfying
the conditions a  given in Table 3. We must also determine  which
sign Eq.(\ref{signa}) or Eq.(\ref{signb}) characterises
$R^{a_1...a_{D-3},a_{D-2}}$ under the action of the conjugated
involution $\Omega^\prime$ given by Eq.(\ref{newinvolve}). Using
Eq.(\ref{kisi}) one has $\mathrm{sign}(\Omega^\prime R^{4...D,D}) =
\mathrm{sign}(\Omega R^{4...D,D}) = -1$ because there is an odd number of time
coordinates in $\{4,...,D-1\}$ before Weyl reflection. The nature
of all these coordinates changes under the action of $W_{\a_D}$.
Therefore if $D$ is even, we have an odd number of time
coordinates in this subset and $\mathrm{sign}(\Omega^\prime
R^{4...D,D})$ satisfies Eq.(\ref{signa}). If $D$ is odd we get an
even number of time coordinates yielding to the other sign
Eq.(\ref{signb}). Therefore, the action of $W_{\alpha_D}$ on the
$\Omega$'s characterised by the signatures of Table \ref{gravity1}
yields $\Omega^\prime$'s characterised by the signatures given in
Table \ref{newsigna}.
\begin{table}[h]
\caption{\small  $\Omega^\prime$'s obtained by the Weyl reflection
$W_{\a_D}$ from $\Omega$'s given in Table \ref{gravity1}}
\begin{center}
\begin{tabular}{|c|l|c|}
\hline &        signature $\Omega'$ & conditions on $\Omega$ \\
\hline a.1. &
            $(S,T,(-)^D)$ & $T \geq 3 $ and $S\geq 3$ \\
       a.2. &
            $(S-4,T+4,(-)^D)$ & $T \geq 3 $ and $S\geq 4$ \\
       a.3. &
            $(S,T,(-)^D)$ & $T \geq 3 $ and $S\geq 4$ \\
       a.4.  &
            $(S,T,(-)^D)$ & $T \geq 3 $ and $S\geq 3$ \\
\hline \end{tabular}
\end{center}
\label{newsigna} \end{table}

\item b. We get the same possibilities as those of case a (see Table \ref{gravity1})  except for the
coordinate $D-1$ which is different. Therefore the new signatures
will be the same. Only the conditions on $S$ and $T$ can differ.
These conditions are given in Table \ref{newsignb}.
\begin{table}[h]
\caption{\small  $\Omega^\prime$'s obtained by the Weyl reflection
$W_{\a_D}$ from $\Omega$'s given in Table \ref{gravity1} with the
nature of the coordinate $D-1$ changed }
\begin{center}
\begin{tabular}{|c|c|c|}
\hline &          signature $\Omega'$ & conditions on $\Omega$ \\
\hline b.1. & $(S,T,(-)^D)$ & $T \geq 3 $ and $S\geq 4$ \\
       b.2. & $(S-4,T+4,(-)^D)$ & $T\geq 1 $ and $S\geq 5$ \\
       b.3. &  $(S,T,(-)^D)$ & $T \geq 3 $ and $S\geq 3$ \\
       b.4. &  $(S,T,(-)^D)$ & $T \geq 5$ and $S\geq 2$ \\
\hline \end{tabular}
\end{center}
\label{newsignb}
\end{table}

\item c. The coordinates $3$ and $4$ are of the same nature,  $D-1$ and $D$ are of different
nature.  Moreover there are an odd number of time coordinates in $\{ 3,5,...,D-1\}$ and therefore an even number of time
coordinates in the complementary subset $\{1,2,4,D\}$, i.e 0, 2 (1
being always space-like in $\g_B$). These conditions lead to the
possibilities given in Table \ref{gravity3}. The new signatures
are given in Table \ref{newsignc} (the sign for
$R^{a_1...a_{D-3},a_{D-4}}$ is determined by applying a similar
reasoning as the one developed in case a).

\begin{table}[h]
\caption{\small $\Omega$'s leading to  signature changes under
the Weyl reflection $W_{\alpha_D}$ in case c. There is an odd number of time coordinates in
the subset $\{4,...,D-1\}$}
\begin{center}
\begin{tabular}{|c|ccc|ccc|c|}
\hline &       1 & 2 & 3 & 4 & ... & D-1 & D \\
\hline c.1. &  s & s & s & s & ... & t & s \\
       c.2. &  s & t & t & t & ... & t & s \\
       c.3. &  s & t & s & s & ... & s & t \\
       c.4. &  s & s & t & t & ... & s & t \\
\hline \end{tabular}
\end{center}
\label{gravity3} \end{table}

\begin{table}[h]
\caption{\small    $\Omega^\prime$'s obtained by the Weyl reflection
$W_{\a_D}$ from $\Omega$'s given in Table \ref{gravity3}}
\begin{center}
\begin{tabular}{|c|c|c|}
\hline &          signature $\Omega'$ & conditions on $\Omega$ \\
\hline c.1. & $(S-4,T+4,(-)^D)$ & $T \geq 1 $ and $S\geq 5$ \\
       c.2. & $(S,T,(-)^D)$ & $T\geq 5 $ and $S\geq 2$ \\
       c.3. &  $(S,T,(-)^D)$ & $T \geq 3 $ and $S\geq 4$ \\
       c.4. &  $(S,T,(-)^D)$ & $T\geq 3$ and $S\geq 3$ \\
\hline \end{tabular}
\end{center}
\label{newsignc}
\end{table}

\item d. We get the same signature as those of case c (see Table \ref{gravity3}) except for
the nature of the coordinate $D-1$. Again, only the conditions on
$T$ and $S$ can differ. The new signatures are given in Table
\ref{newsignd}.

\begin{table}[h]
\caption{\small  $\Omega^\prime$'s obtained by the Weyl reflection
$W_{\a_D}$ from $\Omega$'s given in Table \ref{gravity3} with the
nature of the coordinate $D-1$ changed }
\begin{center}
\begin{tabular}{|c|c|c|}
\hline &          signature $\Omega'$ & conditions on $\Omega$ \\
\hline d.1. & $(S-4,T+4,(-)^D)$ & $T \geq 1 $ and $S\geq 6$ \\
       d.2. & $(S,T,(-)^D)$ & $T \geq 3 $ and $S\geq 3$ \\
       d.3. &  $(S,T,(-)^D)$ & $T \geq 3 $ and $S\geq 3$ \\
       d.4. &  $(S,T,(-)^D)$ & $T \geq 5$ and $S\geq 2$ \\
\hline \end{tabular}
\end{center}
\label{newsignd}
\end{table}

\end{itemize}
\noindent \textbf{Summary}: From a signature $(T,S,+)$ with $T$
odd, we can
 reach the signatures
\bea (S-4,T+4,(-)^D) \ \ & & \mathrm{if} \ \ \{ \ T \geq 3 \
\mathrm{and} \ S \geq
4 \} \ \mathrm{or} \ \{T \geq 1 \ \mathrm{and} \ S \geq 5 \} \nn \\
\label{cong}
(S,T,(-)^D) \ \ & & \mathrm{if} \ \  \{ T\geq 5 \ \mathrm{and} \ S
\geq 2 \} \ \mathrm{or} \ \{ T \geq  3 \ \mathrm{and} \ S \geq 3
\}. \label{cond1} \eea
In order to find the Weyl orbits of $(1,D-1,+)$ we need to distinguish between  $D$ even and $D$ odd.

\textit{\textbf{$D$ even}}: The conditions given by Eq.
(\ref{cond1}) simplify to
\bea (S-4,T+4,+) \ \ & & \mathrm{if} \ \  \{T \geq 1 \ \mathrm{and} \ S \geq 5 \} \nn \\
\label{condsi}(S,T,+) \ \ & & \mathrm{if} \  \ \{ T \geq  3 \ \mathrm{and} \ S
\geq 3 \}. \label{cond2} \eea

If we start from $(1,D-1,+)$,  after the action of  $W_{\a_D}$ one
gets $(D-5,5,+)$ which is of the
generic type $(T,S,+)$ with $T$ odd furthermore all the other
signatures that we reach given by Eq.(\ref{condsi}) are of this
type. Therefore we can use the above analysis  and  taking into
account the conditions Eq.(\ref{condsi}) we conclude that when $D$
is even, the signatures in the $W_{\a_D}$ orbit of $(1,D-1,+)$ for
$\g_B =A_{D-3}^{++}$  are given by ($n$ is an integer)

\bea (1,D-1,+) \nn \\
(1+4n , D-1-4n,+) \  && 3 \leq 4n+1 \leq D-3 \nn \\
\label{condev}
( D-1-4n, 1 + 4n , +) \ &&  5 \leq 4n+1 \leq D-1. \eea

\textit{\textbf{$D$ odd}}: After the action of  $W_{\a_D}$ the
signature $(D-5,5,-)$ is not of the type $(T,S,+)$ with $T$ odd.
To determine the orbit of $(1,D-1,+)$, we have thus to analyse the
signatures of the form $(T,S,-)$ with $T$ even and $S$ odd.

\item \textbf{Step 2}: Let us consider an  involution $\Omega$ characterised
by a signature $(T,S,-)$ where $T$ is
even and $S$ is odd ($D$ is odd). The discussion of the different
possible cases of signature change is similar to the ones
discussed in step 1. Indeed, the  even number
of time coordinates balances the   minus sign
 for the kinetic term of
$R^{a_1...a_{D-3},a_{D-4}}$ (see Eq.(\ref{signb})). We must be careful as far as the
conditions on $T$ and $S$ are concerned. The only difference with
the Tables \ref{newsigna}, \ref{newsignb}, \ref{newsignc},
\ref{newsignd} is that we have for $D$ odd the opposite parity for
the number of times in the coordinates $\{4,...,D-1\}$. Therefore,
starting with $(T,S,-)$ we will reach the signature $(S,T,+)$
under the following conditions, \bea (S,T,+) \ \ & & \mathrm{if} \
\  \{ T\geq 2 \ \mathrm{and} \ S \geq5 \} \ \mathrm{or} \ \{ T
\geq  4 \
\mathrm{and} \ S \geq 3 \}. \label{cond2} \eea
The sign + given by Eq.(\ref{signa})  for $\Omega^\prime
R^{a_1...a_{D-3},a_{D-4}}$ is obtained by a reasoning similar to the one
given below Table \ref{gravity1} taking into account the fact that
the number of dimensions is odd.

The new signature Eq.(\ref{cond2}) is of the type considered in
step 1. The conditions for getting new signatures by acting
again with $W_{\a_D}$ on this signature can thus be deduced from Eq.(\ref{cong}).
Starting from $(1,D-1,+)$ the step 1 gives us
$(D-5,5,-)$, then step 2 gives us $(5,D-5,+)$, step 1 can be used
again to obtain $(D-9,9,-)$. Repeating the argument, all the new
signatures are obtained using ``step 1" or ``step 2".

We conclude that when $D$ is odd, the following signatures can be
reach,
\bea (1+4n , D-1-4n,+) \  && 1 \leq 4n+1 \leq D-2 \nn \\
\label{condodd} ( D-1-4n, 1 + 4n ,-) \ &&  1< 4n+1 \leq D. \eea

\end{itemize}

\noindent We can rewrite Eq.(\ref{condev}) and Eq.(\ref{condodd})
in a more concise way and conclude that for {\it all} $D$ (odd and
even) the signatures of $\g_B$ in the Weyl orbit of $(1,D-1,+)$
are given by

\bea (1+4n , D-1-4n,+) \  && 0 \leq n \leq [ \frac {D-3}{4} ] \nn \\
\label{confin}
( D-1-4n, 1 + 4n ,  (-)^D) \ &&  1< n \leq   [ \frac {D-1}{4} ]  ,
\eea where $n$ is an integer and $[x]$ is the integer part of $x$.
\subsubsection{$D=5$}

The reasoning of the previous section cannot apply for $D=5$
because the conditions on $\Omega$ to obtain new signatures
$\Omega^\prime$ given in Table \ref{con1} assumed $4< D-1$. These
conditions give in this case  $\mathrm{sign} (\Omega K^3{}_4 )\neq
\mathrm{sign}(\Omega R^{35,5})$ \textit{and} $\mathrm{sign}
(\Omega K^4{}_5) \neq \mathrm{sign}(\Omega R^{45,4})$. Starting
from a signature $(1,4,+)$, this implies that the coordinate 4
must be the time. We get by acting with the Weyl reflection
generated by $\alpha_5$ the signature $(0,5,-)$. Consequently all the
possible signatures for $D=5$ are

\bea (1,4,+) \ (0,5,-) .\eea

\subsubsection{$D=4$}

\begin{figure}[h]
\caption{ \small Dynkin diagram of $A_1^{+++}$}
\begin{center}
\scalebox{.8}{
\begin{picture}(180,60)
\put(0,-5){$1$} \put(40,-5){$2$} \put(80,-5){$3$} \put(90,40){$4$}
\thicklines \multiput(0,10)(40,0){3}{\circle{10}}
\multiput(5,10)(40,0){2}{\line(1,0){30}}
\put(80,45){\circle*{10}}
\put(76,15){\line(0,1){30}} \put(79,15){\line(0,1){29}}
\put(82,15){\line(0,1){30}} \put(85,15){\line(0,1){30}}
\end{picture}
}
\end{center}
\label{second}
\end{figure}
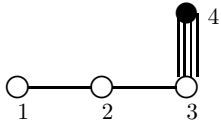

The Dynkin diagram of $A_1^{+++}$ is depicted in Fig.\ref{second}.
To get a signature change due to the action of the Weyl reflection
generated by $\alpha_4$ we need sign($\Omega K^3{}_4$) $\neq$
sign($\Omega R^{3,4}$) which is not possible since we started from
$(1,3,+)$. The symmetric tensor $R^{a,b}$ is the representation
[2,0,0] of $A_3$ that occurs at level  one. The only possible
signature is the Minkowskian one, \bea (1,3,+) .\eea

\subsection{$D_{D-2}^{+++}$}

\subsubsection{$D>6$}
\noindent They are two simple roots not belonging to the gravity line,
namely $\a_D$ and $\a_{D+1}$ (see Fig.\ref{first}). Given a
signature $(T,S,\varepsilon)$, the first sign in the set
$\varepsilon$ is associated to the generator\footnote{The tensor
$R^{a_1a_2}$ is in the representation $[0,1,0,\dots,0,0]$ of
$A_{D-1}$ that occurs at level $(1,0)$, the tensor $R^{a_5...a_D}$
is in the representation $[0,\dots,0,1,0,0,0]$ of $A_{D-1}$ that
occurs at level $(0,1)$ \cite{weke}. The level $(l_1,l_2)$ of a
root $\a$ of $D_{D-2}^{+++}$ can be read in its
 decomposition  in terms of the simple roots $\a = m_1 \a_1+... + m_{D-1} \a_{D-1} + l_1 \a_D + l_2
\a_{D+1}$.  } $R^{a_1a_2}$ and the second one to the generator
$R^{a_5...a_D}$ (see Eqs. (\ref{signa}) and (\ref{signb})). We will analyse the possible signature changes
due to $W_{\a_D}$ and $W_{\a_{D+1}}$.

1) The Weyl reflection generated by the root $\a_{D}$ will never
\textit{non trivially} change the signature. Indeed, its
 action on the simple root $\a_{D-2}$ and the corresponding
action of $U_{W_{\a_D}}$ on the simple generator  $R^{D-2\, D}$ are,

\begin{center}
\begin{tabular}{lrcllrcl}
$W_D$:& $\a_{D-2}$ & $ \leftrightarrow$ & $ \a_{D-2} + \a_{D}$ &
$U_{W_D}$: & $K^{D-2}{}_{D-1}$ & $\leftrightarrow$ & $\sigma
R^{D-2D}.$
\end{tabular} \end{center}

\noindent Therefore to obtain a change of signature one needs (see Eq.(\ref{newinvolve})):
$\mathrm{sign}( \Omega K^{D-2}{}_{D-1}) \neq \mathrm{sign}( \Omega
R^{D-2D})$. We have
$$\mathrm{sign} ( \Omega R^{D-1D} )=
-\mathrm{sign}(\Omega R^{D-2D}) . \mathrm{sign}(\Omega
K^{D-2}{}_{D-1}),$$  thus $\mathrm{sign}(\Omega R^{D-1D}) = +$
implies that the coordinates $D-1$ and $D$ are of different nature because $R^{D-1\, D}$
satisfies Eq.(\ref{signa}). So even if
$\Omega'K^{D-2}{}_{D-1} \neq \Omega K^{D-2}{}_{D-1}$ we will not
have a \textit{non trivial} signature change but just an exchange
between the nature of the coordinates $D-1$ and $D$.

2) We now consider the action of the Weyl reflection generated
by the root $\a_{D+1}$. Its action on $\a_4$ is
\begin{center}
\begin{tabular}{lrcllrcl}
$W_{D+1}$:& $\a_{4}$ & $ \leftrightarrow$ & $ \a_4 + \a_{D+ 1}$ &
$U_{W_{D+1}}$: & $K^4{}_5$ & $\leftrightarrow$ & $ \rho
R^{46...D}.$
\end{tabular} \end{center}

\noindent  In order to have a signature change, we need sign($ \Omega K^4{}_5)
\neq$ sign($ \Omega
 R^{46...D})$. This leads to two possibilities explicitly given in
 Table \ref{conddnD}. The condition \textbf{A} means that the nature of the coordinates
 4 and 5 are different \textit{and} that if we start from $( T
, S, \pm , +)$ (resp. $(T  ,S ,\pm ,-)$) there is an even (resp. odd)
number of times in $\{4,6,...,D\}$ . Whereas condition \textbf{B}
means that that the nature of the coordinates
 4 and 5 are the same \textit{and} that if we start from $( T , S , \pm  ,
+)$ (resp. $(T ,S ,\pm ,-)$) there is an odd (resp. even) number of
times in $\{4,6,...,D\}$.

\begin{table}
\caption{\small  Conditions on $\Omega$'s leading to non-trivial
signature change under the Weyl reflection $W_{\alpha_D}$ for
$S_{D_{D-2}^{++}}$ }
\begin{center}
\begin{tabular}{|c|cc|}
\hline & sign($\Omega K^4{}_5$) & sign($\Omega R^{46...D}$) \\
\hline \textbf{A} &+ & - \\
 \textbf{B}  &- & + \\
\hline
\end{tabular}
\end{center}
\label{conddnD} \end{table}

\begin{itemize} \item By analogy with the $A_{D-3}^{+++}$ case, we start from the signature $(T,S,+,+)$ where $T$ is an odd number.

\textbf{A.} There is an odd number of times in $\{1,2,3,5\}$, i.e
1 or 3. These time coordinates can be distributed as in Table
\ref{newsignAdn}. The new
signatures $\Omega^\prime$, also given in Table \ref{newsignAdn},
are deduced from the fact that the action of the Weyl reflection
generated by $\alpha_{D+1}$ will change the nature of the
coordinates greater than 4.

\begin{table}
\caption{ \small $\Omega$'s leading to  signature changes under
the Weyl reflection $W_{\alpha_{D+1}}$ and the related new signatures
$\Omega^\prime$'s (in the case \textbf{A}).}
\begin{center}
\begin{tabular}{|c|ccccc|c|c|c|}
\hline   & 1 & 2 & 3 & 4 & 5  &$\Omega^\prime$ & $T\geq $ & $S \geq$\\
 \hline i. & s & s& t & t & s  &$(S,T ,+,(-)^D)$& 3 & 3\\
 ii. & s & s & s & s & t  & $(S-4, T + 4,+,(-)^D)$ & 1 & 4\\
 iii. & s & t & t & s & t  &$(S,T ,+,(-)^D)$& 3 & 2\\
\hline \end{tabular}
\end{center}
\label{newsignAdn}
\end{table}

\textbf{B.} There is an even number of times in $\{1,2,3,5\}$ i.e
0, 2 or 4. 4 is excluded here because we want that 1 is
space-like. These time coordinates can be distributed as in Table
\ref{newsignBdn}. The new signatures are also given in Table
\ref{newsignBdn}.

\begin{table}
\caption{$\Omega$'s leading to  signature changes under the Weyl
reflection $W_{\alpha_{D+1}}$ and the related new signatures
$\Omega^\prime$'s (in the case \textbf{B}).}
\begin{center}
\begin{tabular}{|c|ccccc|c|c|c|}
\hline   & 1 & 2 & 3 & 4 & 5  & $\Omega^\prime$ & $T\geq$& $S \geq$\\
\hline i. & s & s & s & s & s  &$(S-4 , T + 4,+,(-)^D)$& 1  & 5\\
 ii. & s & s & t & t & t & $(S, T ,+,(-)^D)$& 3 & 2\\
 iii. & s & t & t & s & s  & $(S, T ,+,(-)^D)$& 3 & 3\\
\hline \end{tabular}
\end{center}
\label{newsignBdn}
\end{table}

\noindent $\Rightarrow$ if $D$ is even, from a signature
$(T,S,+,+)$ we can reach the signatures $(S,T,+,+)$ (if $T \geq 3$
and $S \geq 3$) and $(S-4,T+4,+,+)$ (if $T \geq 1$ and $S \geq 5$
).

\noindent $\Rightarrow$ if $D$ is odd, from a signature
$(T,S,+,+)$ we can reach the signatures $(S,T,+,-)$  (if $T\geq 3
$ and $S \geq 2$) and $(S-4,T+4,+,-)$ (if $T \geq 1$ and $S \geq
4$).

\item From the signature $(T,S,+,-)$ where $T$
is an even number and $S$ odd ($D$ is odd). By the same procedure,
we can conclude that the signatures $(S,T,+,+)$ can be reached (if
$T\geq 2 $ and $S \geq 3$) and the signatures $(S-4,T+4,+,+)$ (if
$S \geq 5$).

\end{itemize}

\noindent \textbf{Summary}: We get the same signatures as the ones
of pure gravity as expected since $W_{\a_D}$ does not change the
signature and only $W_{\a_{D+1}}$ has a non-trivial action,
\bea (1+4n , D-1-4n,+,+) \  && 0 \leq n \leq [ \frac {D-3}{4} ] \nn \\
( D-1-4n, 1 + 4n , +, (-)^D) \ &&  1< n \leq   [ \frac {D-1}{4} ]
, \eea where $[x]$ is the integer part of $x$.
We could have acted with the Weyl reflection generated by the graviphoton lying at level $(1,1)$ instead of acting with $W_{\a_{D+1}}$ and we would have obtained the same results. The sign of the kinetic term of the graviphoton
agrees with Eq.(\ref{confin}).

\subsubsection{$D=6$}
\noindent They are two simple roots out of the gravity line,
namely $\a_6$ and $\a_{7}$ (see Fig.\ref{third}).
\begin{figure}[h]
\caption{ \small Dynkin diagram of $D_4^{+++}$}
\begin{center}
\scalebox{.7}{
\begin{picture}(180,60)
\put(5,-5){1} \put(45,-5){2} \put(85,-5){3} \put(140,-5){4}
\put(165,-5){$5$} \put(140,45){$6$}\put(140,-35){$7$} \thicklines
\multiput(10,10)(40,0){5}{\circle{10}}
\multiput(15,10)(40,0){4}{\line(1,0){30}}
\put(130,50){\circle*{10}} \put(130,15){\line(0,1){30}}
\put(130,-30){\circle*{10}} \put(130,5){\line(0,-1){30}}
\end{picture}
}
\end{center}
\label{third}
\end{figure}
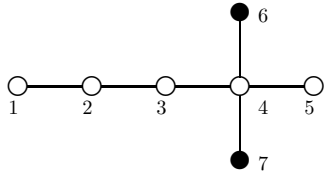 Given a
signature $(T,S,\varepsilon)$, the first sign in the set
$\varepsilon$ is associated to the generator $R^{a_1a_2}$ and the second one to the generator
$\tilde{R}^{a_1a_2}$.

We can immediately conclude that no signature changes are
possible. Indeed the Weyl reflections generated by  the two simple
roots not belonging to the gravity line cannot change the signature in the
same way as the Weyl reflection $W _{\a_D}$ cannot do it in the
previous section.

\subsection{$E_6^{+++}$}

They are two simple roots not belonging to the gravity line,
$\a_8$ and $\a_9$ (see Fig. 1). The non trivial actions of the corresponding Weyl
reflections $W_8$, $W_9$ are

\begin{center}
\begin{tabular}{crclcrcl}
 $W_8$:  & $\a_{5}$ & $ \leftrightarrow$ & $ \a_5 + \a_{8}$& $U_{W_8}$: & $K^5{}_6$ &
$\leftrightarrow$ & $ \rho R^{578}$  \\
         & $\a_{9}$ & $ \leftrightarrow$ & $ \a_9 + \a_{8}$&  & $R$ &
$\leftrightarrow$ & $ \sigma \tilde{R}^{678}$ \\
 $W_9$:  & $\a_{8}$ & $ \leftrightarrow$ & $ \a_8 + \a_{9}$& $U_{W_9}$: & $ R^{678}$&
$\leftrightarrow$ & $ \delta \tilde{R}^{678}$.   \end{tabular}
\end{center}
The tensor  $R^{abc}$ is the representation
$[0,0,1,0,0,0,0]$ of $A_{7}$ that occurs at level\footnote{The
level $(l_1,l_2)$ of a root $\a$ of $E_6^{+++}$ can be read in its
 decomposition  in terms of the simple roots $\a = m_1 \a_1+... + m_7 \a_7 + l_1 \a_8 + l_2
\a_9$. } $(1,0)$, the tensor $R$ is the representation
[0,0,0,0,0,0,0] of $A_7$ that occurs at level $(0,1)$ and $
\tilde{R}^{abc}$ is the representation $[0,0,1,0,0,0,0]$ of $A_7$
that occurs at level $(1,1)$  \cite{weke}. The sign of the kinetic term of
$\tilde{R}^{abc}$ can be deduced from the ones of $R^{abc}$ and
$R$, it is the product of these two signs.

The Weyl reflection generated by $\a_9$ exchanges the signs of
$R^{abc}$ and $\tilde{R}^{abc}$.

To obtain a signature change from $(1,7,+,+)$\footnote{The first
sign characterises  $R^{abc}$ and the second one characterises $R$ (see Eqs.(\ref{signa}) and (\ref{signb})).}, we must act with
the Weyl reflection generated by $\alpha_8$ and there  must be an
odd number of time coordinates in the following subset: 6,7,8. In
all of this cases, the new signature is $(2,6,-,-)$. Now we can
start from this new signature and act with the Weyl reflection
generated by $\alpha_9$ to get the signature $(2,6,+,-)$ or with
the one generated by $\alpha_8$ to get $(5,3,+,+)$. Acting again with the Weyl reflections generated
by  $\alpha_8$ (vertical arrows) and $\alpha_9$ (horizontal arrows) on
these signatures,  we can conclude that all the signatures are

\begin{center}
\begin{tabular}{ccc}
$(1,7,+,+)$ & & \\
$\downarrow$ & & \\
$(2,6,-,-)$ & $\rightarrow$ & $(2,6,+,-)$ \\
$\downarrow$ & & $\downarrow$\\
$(5,3,+,+)$ & & $(3,5,-,+)$ \\
$\downarrow$ & & $\downarrow$\\
$(6,2,-,-)$ & $\rightarrow$ & $(6,2,+,-).$ \\
\end{tabular}
\end{center}
Theses signatures are the expected ones from the gravity, i.e.
$(1,7,+,+)$, $(3,5,-,+)$ and $(5,3,+,+)$, plus new ones.

\subsection{$E_7^{+++}$}

They are two simple roots not belonging to the gravity line, $\a_9$ and
$\a_{10}$. The non trivial actions of the Weyl reflections $W_9$,
$W_{10}$ are

\begin{center}
\begin{tabular}{rrclrrcl}
 $W_9$:  & $\a_{8}$ & $ \leftrightarrow$ & $ \a_8 + \a_{9}$& $U_{W_9}$: & $K^8{}_9$ &
$\leftrightarrow$ & $ \rho R^{8}$  \\
 $W_{10}$:  & $\a_{6}$ & $ \leftrightarrow$ & $ \a_6 + \a_{10}$& $U_{W_{10}}$: & $ K^{6}{}_7$&
$\leftrightarrow$ & $ \delta R^{689}.$   \end{tabular}
\end{center} The tensor  $R^{a}$ is the representation
$[1,0,0,0,0,0,0,0]$ of $A_{8}$ that occurs at level\footnote{The
level $(l_1,l_2)$ of a root $\a$ of $E_7^{+++}$ can be read in its
 decomposition  in terms of the simple roots $\a = m_1 \a_1+... + m_7 \a_7 + m_8 \a_8 +
 l_1
\a_9 + l_2 \a_{10}$. }  $(1,0)$, the tensor $R^{abc}$ is the
representation [0,0,1,0,0,0,0,0] of $A_8$ that occurs at level
$(0,1)$.

 The Weyl reflection generated by the
root $\alpha_9$ will change the signature if  $\mathrm{sign}
(\Omega K^8{}_9) \neq \mathrm{sign}(\Omega R^8)$. The Weyl
reflection generated by the root $\alpha_{10}$ will change the
signature if $\mathrm{sign} (\Omega K^6{}_7) \neq
\mathrm{sign}(\Omega R^{689})$. With these rules we get the
following signatures (a horizontal arrow represents the action of
the Weyl reflection generated by $\alpha_{10}$ and a vertical one
the action of the Weyl reflection generated by $\a_9$),

\begin{center}
\begin{tabular}{cccccccc}
$(1,8,+,+)$ & $\rightarrow$ & $(2,7,-,-)$ & $\rightarrow$ &
$(5,4,+,+)$ & $\rightarrow$& $(6,3,-,-)$\\
$\downarrow$ & & $\downarrow$ & & $\downarrow$ & & $\downarrow$ \\
$(0,9,-,-)$ & $\rightarrow$ & $(3,6,+,+)$ & $\rightarrow$ &
$(4,5,-,-)$ & $\rightarrow$& $(7,2,+,+).$ &
\end{tabular}
\end{center} The signs
in the above signatures refer to the kinetic terms of $R^a$ and
$R^{abc}$ (see Eqs.(\ref{signa}) and (\ref{signb})).

\subsection{$E_8^{+++}$}

The signatures are given in \cite{ehh}. To be complete, we recall
them here, \bea (1,10,+) \  \rightarrow \ (2,9,-) \ \rightarrow \
(5,6,+) \ \rightarrow (6,5,-) \ \rightarrow (9,2,+). \eea The sign
refers to the sign of the kinetic term of $R^{abc}$ which is the
representation [0,0,1,0,0,0,0,0,0,0] of $A_{10}$ that occurs at
level 1.

\section{Non simply laced algebras}
\subsection{$B_{D-2}^{+++}$}

The only non trivial action of the Weyl reflection generated by the short root
$\a_D$ is

\begin{center}
\begin{tabular}{rrclrrcl}
 $W_D$:  $\a_{D-1}$ & $\leftrightarrow$ & $\a_{D-1} + 2
\a_D$ & $U_{W_D}$: $K^{D-1}{}_D$ & $\leftrightarrow$ & $\sigma
R^{D-1D}.$   \end{tabular}
\end{center} The tensor  $R^{ab}$ is the representation
$[0,1,0,,\dots,0]$ of $A_{D-1}$ that occurs at
level\footnote{The level $(l_1,l_2)$ of a root $\a$ of
$B_{D-2}^{+++}$ can be read in its
 decomposition  in terms of the simple roots $\a = m_1 \a_1+...  + m_{D-1} \a_{D-1} +
 l_1
\a_D + l_2 \a_{D+1}$. } $(2,0)$ \cite{weke}. To get a signature
change with this Weyl reflection we need $\mathrm{sign}( \Omega
K^{D-1}{}_D ) \neq \mathrm{sign} ( \Omega R^{D-1D}) $ which is
impossible since the sign of the kinetic term of $R^{ab}$ is given by Eq.(\ref{signa}).

We are therefore left with one simple long root, namely $\a_{D+1}$. The Weyl reflection $W_{\a_{D+1}}$ will
clearly produce the same signature changes as $W_{\a_{D+1}}$ does
for $D_{D-2}^{+++}$. Therefore all possible signatures are the
ones found for $D_{D-2}^{+++}$, i.e the signatures of pure
gravity.
\bea (1+4n , D-1-4n,+,+) \  && 0 \leq n \leq [ \frac {D-3}{4} ] \nn \\
( D-1-4n, 1 + 4n , +, (-)^D) \ &&  1< n \leq   [ \frac {D-1}{4} ]
, \eea where $[x]$ is the integer part of $x$. The first sign
refers to the kinetic term of $R^{a}$ and the second to the one of
$R^{a_5...a_D}$. $R^a$ is the representation [1,0,...,0] of
$A_{D-1}$ that occurs at level $(1,0)$ and $R^{a_5...a_{D}}$ the
representation [0,...,0,1,0,0,0] that occurs at level (0,1)
\cite{weke}.

\subsection{$C_{q+1}^{+++}$}

Clearly, if the Weyl reflections associated to $\a_4$ and $\a_5$
do not change the signature, there will be no signature changes.
Let us first have a look at $\a_5$,
\begin{center}
\begin{tabular}{crclcrcl}
 $W_5$:  & $\a_{4}$ & $\leftrightarrow$ & $\a_{4} +
\a_5$ & $U_{W_5}$: & $R^4$ & $\leftrightarrow$ & $\rho \tilde{R}^4$  \\
         & $\a_{6}$ &  $\leftrightarrow$ & $\a_6+\a_5$ & &  $\tilde{R}$ &
$\leftrightarrow$ & $ \sigma \tilde{\tilde{R}}.$   \end{tabular}
\end{center}  The tensor  $R^{a}$ is the representation
$[1,0,0]$ of $A_{3}$ that occurs at level\footnote{The level
$(l_{q+1},l_q,...,l_1)$ of a root $\a$ of $C_{q-1}^{+++}$ can be
read in its
 decomposition  in terms of the simple roots $\a = m_1 \a_1+ m_2 \a_2+ m_3 \a_3 + l_{q+1} \a_4 +
 ...+l_1
\a_{q+4}$. } $(1,0,...,0)$, the tensor $\tilde{R}^a$ is the
representation [1,0,0] of $A_3$ that occurs at level
$(1,1,0,...,0)$, the tensor $\tilde{R}$ is the representation
$[0,0,0]$ of $A_3$ that occurs at level (0,0,1,0,...,0) and $
\tilde{\tilde{R}}$ is the representation $[0,0,0]$ of $A_3$ that
occurs at level $(0,1,1,0,...,0)$. Because we started with a
signature where all the generators have the sign of their kinetic
term positive, we will not reach new signatures (i.e, here meaning
new signs for the kinetic term of $R^a$ or $\tilde{R}$) with this
reflection.

Let us now look at $\a_4$,

\begin{center}
\begin{tabular}{crclcrcl}
 $W_4$:  &  $\a_{5}$ & $\leftrightarrow$ & $\a_{4} +
\a_5$  & $U_{W_4}$: &  $R$ & $\leftrightarrow$ & $\tilde{R}^4$  \\
         &$\a_{3}$ &  $\leftrightarrow$ & $\a_3+2\a_4$ & &   $K^3{}_4$ &
$\leftrightarrow$ & $R^{34}.$  \end{tabular}
\end{center}  The tensor $R$ is the representation [0,0,0] of $A_3$ that
occurs at level (0,1,0,...,0), $R^{ab}$ is the representation
[0,1,0] of $A_3$ that occurs at level (2,0,...,0). To get a
signature change we need $ \Omega (K^3{}_4) \neq \Omega (R^{34})$
which is impossible because we have Eq.(\ref{signa}) for $R^{34}$.

\subsection{$F_4^{+++}$}

There are two short simple roots not belonging to  the gravity line, namely $\a_6$
and $\a_7$. The generators associated to these simple roots are
given in Table \ref{f4} respectively at level\footnote{The level
$(l_1,l_2)$ of a root $\a$ of $F_4^{+++}$ can be read in its
 decomposition  in terms of the simple roots $\a = m_1 \a_1+... + m_5 \a_5 +
 l_1
\a_6 + l_2 \a_{7}$. } (1,0) and (0,1) \cite{weke}.
\begin{table}
\caption{\small Level decomposition of $F_4^{+++}$  }
\begin{center}
\begin{tabular}{|ccc|}
\hline
$(l_1,l_2)$&$A_5$ weight&Tensor\\
\hline (0,1)&[0,0,0,0,0]&$R$\\
(1,0)&[1,0,0,0,0]&$R^{a}$\\
(1,1)&[1,0,0,0,0]&$\tilde{R}^{a}$\\
(2,0)&[0,1,0,0,0]&$R^{ab}$\\
\hline
\end{tabular}
\end{center}
\label{f4} \end{table} The non trivial action of the Weyl
reflection $W_{\a_6}$ (and of the corresponding conjugaison by a
group element $U_{W_{\a_6}}$) on the simple roots (and on the
corresponding simple generators) are,

\begin{center}
\begin{tabular}{crclcrcl}
 $W_6$:  &  $\a_{5}$ & $\leftrightarrow$ & $\a_{5} +2
\a_6$  & $U_{W_6}$: &  $K^5{}_6$ & $\leftrightarrow$ & $R^{56}$  \\
         &$\a_{7}$ &  $\leftrightarrow$ & $\a_7+\a_6$ & &   $R$ &
$\leftrightarrow$ & $\tilde{R}^{6}.$  \end{tabular}
\end{center}
$W_{\a_6}$ can not change the signature because it needs sign($
\Omega K^5{}_6) \neq$ sign$(\Omega R^{56}$) which is
impossible since the sign of the kinetic term for $R^{ab}$
is always positive. Moreover $W_{\a_6}$ can neither change the
sign of the kinetic term of $R$ because we start from a signature
such that all kinetic terms are characterised by Eq.(\ref{signa}). The action of   $W_{\a_7}$
on the simple roots is,
\begin{center}
\begin{tabular}{crclcrcl}
 $W_7$:  &  $\a_{6}$ & $\leftrightarrow$ & $\a_{6} +
\a_7$  & $U_{W_7}$: &  $R^6$ & $\leftrightarrow$ & $\tilde{R}^{6}.$
\end{tabular}
\end{center}
This reflection can only \textit{a priori} change the sign of the
kinetic term for $R^a$. This sign can change if sign($\Omega R^6$)
$\neq $ sign($\Omega \tilde{R}^6$) which is impossible since we
started from a signature with all the signs given by Eq.(\ref{signa}).
Therefore the only possible signature is \bea (1,5,+,+), \eea where
the first sign refers to  $R^a$ and the second
to $R$.

\subsection{$G_2^{+++}$}

There is only one simple root not belonging to the gravity line, namely
$\a_5$.
\begin{table}
\caption{\small Level decomposition of $G_2^{+++}$}
\begin{center}
\begin{tabular}{|ccc|}
\hline $l$&$A_4$ weight&Tensor\\
\hline 1 & [1,0,0,0] & $R^a$ \\
3&[1,1,0,0]&$\tilde{R}^{abc}$\\
\hline \end{tabular}
\end{center}
\label{g2} \end{table} The only non trivial action of the Weyl
reflection generated by $\a_5$ on the simple roots is (see Table 13)

\begin{center}
\begin{tabular}{lrcllrcl}  $W_5$:$\a_{4}$ & $\leftrightarrow$ & $\a_{4} +3
\a_5$ & $U_{W_5}$: & $K^4{}_5$ & $\leftrightarrow$ &
$\tilde{R}^{455}.$
\end{tabular}
\end{center}
To get a signature change we need sign($ \Omega K^4{}_5$) $\neq$
sign($\Omega (\tilde{R}^{455}$), i.e. 5 is a time coordinate
(because we start with the sign of
$\tilde{R}^{abc}$ given by Eq.(\ref{signa})). The new signature is the euclidian one
$(0,5,-)$. These are the only  signatures  we can reach in agreement with  pure gravity in $D=5$,

\bea (1,4,+) \ \ (0,5,-), \eea where the sign refers to the kinetic
term of $R^a$.
\section*{Acknowledgments}

One of us (L.H.) would like to warmly thank Fran\c{c}ois Englert
for many enjoyable and fruitful conversations. We are grateful to Marc Henneaux and
Arjan Keurentjes for interesting comments. This work was supported in part
by the NATO grant PST.CLG.979008, by the ``Interuniversity
Attraction Poles Programme -- Belgian Science Policy '', by
IISN-Belgium (convention 4.4505.86)  and by the European
Commission FP6 programme MRTN-CT-2004-005104, in which S.~D.,
L.~H. and N.~T. are associated to the V.U.Brussel (Belgium).

\end{document}